\documentclass{article}
\usepackage{graphicx} 
\usepackage{amsmath}
\usepackage{amsfonts}
\usepackage{float}

\title{A MODEL AND PACKAGE FOR GERMAN COLBERT}
\author{Thuong Dang, Qiqi Chen}
\date{April 2025}

\begin{document}

\maketitle

\begin{abstract}
Retrieval-Augmented Generation (RAG) has become a powerful paradigm in industry. A RAG system is a combination of large language models (LLMs) and information retrieval (IR) techniques. While the retrieval component plays a crucial role in the overall performance of RAG pipelines, most research and tools are primarily focused on English-centric datasets and models.

In this work, we introduce a German version for ColBERT, a late interaction multi-dense vector retrieval method, with a focus on RAG applications. We also present the main features of our package for ColBERT models, supporting both retrieval and fine-tuning workflows.
\end{abstract}

\section{Introduction}

The original ColBERT model was proposed by Khattab and Zaharia \cite{colbert}, introducing the MaxSim scoring function based on token-level interactions. The model was trained using a softmax cross-entropy loss over triplets derived from the MS MARCO Ranking \cite{msmarco} and TREC Complex Answer Retrieval (TREC CAR) \cite{trec} datasets, leveraging the English BERT model \cite{bert} as its backbone encoder. The ColBERT MaxSim score can be interpreted as a substitute for the BM25 score used in full-text search; consequently, there are similarities between the ColBERT retrieval method and BM25-based full-text search. This will be discussed in detail in Section \ref{preliminaries}. ColBERT is flexible, and can be used as a first retrieval method or a reranker. ColBERT score is computed on the token similarity level, and can be applied in contexts where keyword similarities are significant. 

ColBERT model was also trained for Japanese \cite{jacolbert} where the author also discussed different strategies to choose hard negatives using multilingual e5 embedding model and BM25. In another paper \cite{jinacolbert}, the authors experimented to compress the output dimension of XLM-RoBERTa with two phases of training. The first phase focused on pairwise training for semantic similarity, whereas the second phase utilized triplet training with multilingual cross-encoder supervision. This second phase is analogous to the training objective of ColBERT v2 \cite{colbertv2}.

In this paper, we introduce a German ColBERT model trained using the original hidden dimensions of German BERT \cite{gbert}. We employ the MS MARCO Passage Ranking dataset with random negative sampling to emphasize recall—a key objective for retrieval-augmented generation (RAG) applications. In addition to the model, we released a ColBERT package that supports embedding, indexing, re-ranking, training, and negative selection methods.

\section{Preliminaries}\label{preliminaries}

In this section, we will briefly review the ColBERT score, its training objective, and its application to retrieval tasks.

\subsection{ColBERT MaxSim Score}\label{maxsim}

The ColBERT MaxSim scores serve as the backbone of the late interaction mechanism. In models such as cross-encoders, a query and a candidate document interact early because they are concatenated into a single sequence. However, as a reranker, ColBERT does not require early interation between query and candidate document, hence, it is more efficient in terms of speed. 

Let $q$ and $d$ be a query and a document, respectively. Let $[q_1\cdots q_m]$ be the embeddings of tokens of $q$, and $[d_1\cdots d_n]$ be the embeddings of tokens of $d$. The \textit{MaxSim score} is defined as
$$
\operatorname{MaxSim}(q, d) = \sum_{i=1}^m \max_{j} S(q_i, d_j),
$$
where $S(q_i, d_j)$ represents the cosine similarity between the tokens $q_i$ and $d_j$. Intuitively, this can be seen as a soft extension of keyword matching: if we define $S(q_i, d_j) = 1$ when $q_i = d_j$ and 0 otherwise, then the MaxSim score reduces to a keyword‑matching ranking function. Consequently, the more exact keyword matches, the higher a document’s rank will be. In ColBERT, this binary match is replaced by semantic similarity, allowing tokens with similar meanings to contribute to the relevance score.

\subsection{Training objective}\label{training}

ColBERT is built upon the base BERT model \cite{bert}. The training objective for ColBERT is as follows. Given a triplet $[q, d^+, d^-]$, where $q$ is a query, $d^+$ a positive document, and $d^-$ a negative document, the loss function is defined as a pairwise softmax cross-entropy loss
\begin{equation}
L = -\log \frac{\exp \operatorname{MaxSim}(q, d^+)}{\exp \operatorname{MaxSim}(q, d^+) + \exp \operatorname{MaxSim}(q, d^-)}
\label{eq:colBERTloss}
\end{equation}

\subsection{Retrieving method}\label{retreival}

The retrieval method of ColBERT is analogous to keyword-based retrieval.  In traditional keyword retrieval, for each keyword $q_i$ in the query, we collect documents that contain $q_i$, then compute BM25 scores between the query and the set of candidate documents, and finally sort the scores for the final ranking. With ColBERT, for each token $q_i$ in the query, we search for the top-$k'$ most similar tokens, collect the corresponding candidate documents, compute MaxSim scores and finally sort the scores to produce the final ranking. Here is an example for the retrieval method using our German ColBERT model.\\

\textbf{Example.} Assume we have two documents
\begin{enumerate}
    \item Paris ist die Hauptstadt von Frankreich.
    \item Berlin ist die Hauptstadt von Deutschland.
\end{enumerate}
After applying the BERT tokenizer, the tokens for all documents appear as
\begin{enumerate}
    \item ['[CLS]', 'Paris', 'ist', 'die', 'Hauptstadt', 'von', 'Frankreich', '.', '[SEP]']
    \item ['[CLS]', 'Berlin', 'ist', 'die', 'Hauptstadt', 'von', 'Deutschland', '.', '[SEP]']
\end{enumerate}

We can embed all tokens using BERT and then store the embeddings in a flat FAISS index. Let our query be 'Was ist die Hauptstadt von Frankreich?', then query tokens appear as

['[CLS]', 'Was', 'ist', 'die', 'Hauptstadt', 'von', 'Frankreich', '?', '[SEP]']. \\
Retrieving with FAISS on the query tokens for the top 2 results, we obtain the index array

\[
I = \begin{bmatrix}
7 & 5 \\
7 & 5 \\
7 & 3 \\
3 & 7 \\
4 & 3 \\
5 & 7 \\
6 & 5 \\
7 & 5 \\
8 & 7 \\
\end{bmatrix}.
\]

In the similarity matrix $I$, the fifth token in the query, "Hauptstadt", corresponds to the fifth row. It has the highest similarity with the fourth and third tokens in the document (counting from zero), namely "Hauptstadt" and "die". We also obtain the distance array from FAISS retrieval

\[
D = \begin{bmatrix}
0.9520 & 0.9514 \\
0.9568 & 0.9556 \\
0.9563 & 0.9547 \\
0.9604 & 0.9480 \\
0.9642 & 0.9090 \\
0.9755 & 0.9371 \\
0.9682 & 0.9101 \\
0.9594 & 0.9586 \\
0.9471 & 0.8611 \\
\end{bmatrix}.
\]

In the distance matrix $D$, the fifth token "Hauptstadt" has cosine similarity 0.9642 and 0.9090 to the fourth and third tokens in the document. Identifying the most similar tokens, we map them back to their original documents and compute the ColBERT score. As a result, the final top 1 result is:

\begin{itemize}
    \item Document ID: 0, Score: 8.6399, Text: \textit{Paris ist die Hauptstadt von Frankreich.}
\end{itemize}

We can further normalize the score by dividing the ColBERT scores by the number of tokens in the query.

\subsection{Indexing method}\label{indexing}

Based on the retrieval method above, the indexing method of ColBERT can be described as follows. Using the BERT tokenizer, we can tokenize documents into tokens, and compute token embeddings with the trained ColBERT. To facilitate the traceability of each token's vector representation to its corresponding document, we establish a mapping between tokens and documents.

Referring back to the example in Section \ref{retreival}, the mapping of tokens to their respective documents can be represented as the array below

\begin{table}[ht]
\centering
\begin{tabular}{|c|c|}
\hline
\textbf{Token} & \textbf{Document ID}\\
\hline
'[CLS]'  & 1 \\
'Paris' & 1 \\
'ist' & 1 \\
$\cdots$ & $\cdots$ \\
'[SEP]' & 1 \\
'[CLS]'  & 2 \\
'Berlin' & 2 \\
'ist' & 2 \\
$\cdots$ & $\cdots$ \\
'[SEP]' & 2 \\
\hline
\end{tabular}
\caption{Token to doc map array}
\label{tab:token2doc}
\end{table}

Subsequently, we index the embeddings to enable vector-similarity searches, utilizing tools such as the FAISS library. Because ColBERT indexes on the token level, when the dataset scales, it can be problematic to use a flat index, since the search time grows linearly as \( O(Nd) \), with \( N \) the number of vectors in database of dimension \( d \). In practice, when the dataset is large and deletions are infrequent, an HNSW index \cite{HNSW} can be employed. It offers a search time complexity of approximately \( O(d \log N) \). Another option is an IVF-PQ index \cite{IVF}, which reduces the search space and has a search time complexity of \( O(kd + (N/k)d) \), with \( k \) the number of Voronoi cells. 

\subsection{Re-ranking method}
In addition to functioning as a retrieval model (as described in Section \ref{retreival}), ColBERT can also serve as a reranker for other retrieval methods \cite{colbert}. Re-ranking with ColBERT can be performed in two ways:

\begin{itemize}
    \item Batch re-ranking without pre-saving an index: Compute the token embeddings for the query and each candidate document separately, then rank the candidate documents based on their MaxSim scores with respect to the query embeddings. This approach differs from models such as cross-encoders, which require a concatenation between the query and each candidate document and rely on the contextualized [CLS] token for score calculation. Although cross-encoder rerankers are powerful, their scores lack interpretability, making it difficult to establish a threshold for filtering results.

    \item Pre-indexing approach: For datasets of moderate size, pre-computed indices can be saved to disk, eliminating the need for repeated tokenization and embedding computations. Each document $d$ is represented as a matrix of its token embeddings $[d_1\cdots d_n]$. For a given query $q$, compute its token embeddings $[q_1\cdots q_m]$ and then calculate the MaxSim score between $q$ and each candidate documents $d$ as described in Section \ref{maxsim}. The documents are then ranked based on their MaxSim scores, and the top $k$ documents are selected.
\end{itemize}

\section{German ColBERT}
In this section, we describe our German ColBERT model. It was trained on the MS MARCO Passage Ranking dataset, translated using the fairseq-wmt19-en-de model \cite{transmodel}. Our strategy focuses on maximizing recall for RAG applications; hence, we employ random negative sampling. The model was trained on 3.4 million triplets, and its training objective follows that of the original ColBERT as described ~\eqref{eq:colBERTloss} in Section \ref{training}.

The tables below compare recall and NDCG scores between BM25 and our German ColBERT (using top-$k'$ nearest tokens = 100) on two datasets: the antique dataset \cite{antique}, translated using the model in \cite{transmodel}, and the miracl-de-dev dataset \cite{miracl}. The antique dataset includes approximately 3100 queries, while the miracl-de-dev dataset contains around 300 queries.

\begin{table}[htbp]
\centering
\begin{tabular}{|l|c|c|c|c|c|}
\hline
\textbf{Metric} & \textbf{@1} & \textbf{@5} & \textbf{@10} & \textbf{@20} & \textbf{@50} \\
\hline
\multicolumn{6}{|c|}{\textbf{BM25}} \\
\hline
Recall  & 0.1144 & 0.4500 & 0.6797 & 0.7771 & 0.8445 \\
NDCG    & 0.2984 & 0.3731 & 0.4656 & 0.4990 & 0.5166 \\
\hline
\multicolumn{6}{|c|}{\textbf{Our German ColBERT}} \\
\hline
Recall  & 0.1772 & 0.6376 & 0.8710 & 0.9197 & 0.9457 \\
NDCG    & 0.3836 & 0.5262 & 0.6204 & 0.6382 & 0.6448 \\
\hline
\end{tabular}
\caption{Evaluation of BM25 and ColBERT on the \textit{miracl-de-dev} dataset across different cutoff values $k$.}
\label{tab:ir-metrics}
\end{table}

\pagebreak

\begin{table}[htbp]
\centering
\begin{tabular}{|l|c|c|c|c|c|}
\hline
\textbf{Metric} & \textbf{@1} & \textbf{@5} & \textbf{@10} & \textbf{@20} & \textbf{@50} \\
\hline
\multicolumn{6}{|c|}{\textbf{BM25}} \\
\hline
Recall  & 0.0504 & 0.1332 & 0.1699 & 0.2079 & 0.2590 \\
NDCG    & 0.3503 & 0.2584 & 0.2367 & 0.2385 & 0.2567 \\
\hline
\multicolumn{6}{|c|}{\textbf{Our German ColBERT}} \\
\hline
Recall  & 0.0602 & 0.1798 & 0.2410 & 0.2996 & 0.3806 \\
NDCG    & 0.4471 & 0.3484 & 0.3272 & 0.3331 & 0.3619 \\
\hline
\end{tabular}
\caption{Evaluation of BM25 and ColBERT on the translated \textit{antique} dataset across different cutoff values $k$.}
\label{tab:ir-metrics-antique}
\end{table}

Using the two test datasets, we also computed evaluation metrics by first retrieving the top 100 results with BM25 and  subsequently re-ranking them using our German ColBERT model.

\begin{table}[htbp]
\centering
\begin{tabular}{|l|c|c|c|c|c|}
\hline
\textbf{Metric} & \textbf{@1} & \textbf{@5} & \textbf{@10} & \textbf{@20} & \textbf{@50} \\
\hline
Recall  & 0.1840 & 0.5930 & 0.8103 & 0.8588 & 0.8656 \\
NDCG    & 0.4066 & 0.5127 & 0.6002 & 0.6165 & 0.6186 \\
\hline
\end{tabular}
\caption{Evaluation on the \textit{miracl-de-dev} dataset using BM25 to retrieve top 100 candidates, reranked by our German ColBERT model.}
\label{tab:rerank-miracl}
\end{table}

\begin{table}[htbp]
\centering
\begin{tabular}{|l|c|c|c|c|c|}
\hline
\textbf{Metric} & \textbf{@1} & \textbf{@5} & \textbf{@10} & \textbf{@20} & \textbf{@50} \\
\hline
Recall  & 0.0642 & 0.1706 & 0.2109 & 0.2456 & 0.2821 \\
NDCG    & 0.4561 & 0.3369 & 0.3045 & 0.2993 & 0.3099 \\
\hline
\end{tabular}
\caption{Evaluation on the translated \textit{antique} dataset using BM25 to retrieve top 100 candidates, reranked by our German ColBERT model.}
\label{tab:rerank-antique}
\end{table}

\section{Our package for ColBERT}
Along with the German ColBERT model, we also release a package for ColBERT \cite{colbertkit} that includes the following main features:

\begin{itemize}
\item The package provides an indexing method. The primary parameters include the \textit{token-to-document mapping} array as discussed in section \ref{indexing} and the \textit{token embeddings} of the documents. To retrieve, we first load the FAISS index of token embeddings and the token to doc map array. The process then involves embedding the query tokens, retrieving the top-$k'$ nearest tokens, mapping these back to candidate documents using the mapping array, computing the MaxSim scores, and finally sorting them to obtain the top $k$ results.

\item Re-ranking can be accelerated by batching on GPUs. The primary inputs for this include a \textit{query} and a list of \textit{document candidates}.

\item A training script is provided that supports training from checkpoints. The primary input is a triplet dataset containing three columns: sentence, positive sentence, and negative sentence; other parameters are standard. 

\item Evaluation scripts are included to compute key IR metrics such as NDCG and recall for both ColBERT and BM25, supporting both embedding-based retrieval and re-ranking methods. More detail can be found in our documentation for the package \cite{colbertkitdoc}. 

\item Recognizing that collecting negatives in real-world applications can be challenging, we provided scripts to choose negatives from a dataset with two strategies

\begin{itemize}
\item Random negatives: primary input is a data frame with two columns \textit{sentence} and \textit{positive sentence}. A random negative will be randomly selected from other positive sentences.

\item Hard negatives: An additional sentence transformer embedding model is employed to retrieve negatives that are similar to the query.
\end{itemize}    

The output is a triplet dataset with three columns: \textit{sentence}, \textit{positive sentence} and \textit{negative sentence}, that is ready for training or fine-tuning.
\end{itemize}

\bibliographystyle{plain} 
\bibliography{references} 

\end{document}